\documentclass{sig-alternate}
\usepackage{mathptmx} %

\usepackage[normalem]{ulem}
\usepackage[hyphens]{url}
\usepackage[sort,nocompress]{cite}
\usepackage[final]{microtype}
\usepackage[keeplastbox]{flushend}

\usepackage{amsmath,amssymb,amsfonts}
\usepackage{textcomp}
\usepackage{xcolor}
\usepackage{subfig}
\usepackage{graphicx}
\usepackage{algorithmic}
\usepackage[ruled,vlined]{algorithm2e}
\usepackage[flushleft]{threeparttable}
\usepackage{comment}
\usepackage{multirow}
\usepackage{siunitx} %
\usepackage{booktabs}
\usepackage{enumitem}
\def\BibTeX{{\rm B\kern-.05em{\sc i\kern-.025em b}\kern-.08em
    T\kern-.1667em\lower.7ex\hbox{E}\kern-.125emX}}

\usepackage[bookmarks=true,breaklinks=true,letterpaper=true,colorlinks,citecolor=blue,linkcolor=blue,urlcolor=blue]{hyperref}

\newcommand{\nameofwork}{MNF}

{
}

\title{Multiply-and-Fire (MNF): An Event-driven Sparse Neural Network Accelerator \vspace{-1.0cm} }
\numberofauthors{4}
\author{
\alignauthor
Miao Yu\titlenote{} \\
    \affaddr{National University of Singapore} \global\titlenotecount 0\relax %
\alignauthor
Tingting Xiang\titlenote{These authors contributed equally to this work.} \\
    \affaddr{National University of Singapore}
\alignauthor
Venkata Pavan Kumar Miriyala \\
    \affaddr{National University of Singapore}
\and
\alignauthor
Trevor E. Carlson \\
    \affaddr{National University of Singapore}
}

\begin{document}
\maketitle
\pagestyle{plain}

\begin{abstract}

Machine learning, particularly deep neural network inference, has become a vital workload for many computing systems, from data centers and HPC systems to edge-based computing. As advances in sparsity have helped improve the efficiency of AI acceleration, there is a continued need for improved system efficiency for both high-performance and system-level acceleration.

This work takes a unique look at sparsity with an event (or activation-driven) approach to ANN acceleration that aims to minimize useless work, improve utilization, and increase performance and energy efficiency. Our analytical and experimental results show that this event-driven solution presents a new direction to enable highly efficient AI inference for both CNN and MLP workloads.

This work demonstrates state-of-the-art energy efficiency and performance centring on activation-based sparsity and a highly-parallel dataflow method that improves the overall functional unit utilization (at 30 fps). This work enhances energy efficiency over a state-of-the-art solution by 1.46$\times$. Taken together, this methodology presents a novel, new direction to achieve high-efficiency, high-performance designs for next-generation AI acceleration platforms.

\end{abstract}

\section{Introduction}

In recent years, convolutional neural networks (CNNs) have seen significant classification accuracy improvements. They have become the key approach in solving tasks such as object detection, image classification and analysis and semantic segmentation~\cite{deeplearning}. However, these deep networks~\cite {10.1145/3065386, DBLP:journals/corr/SimonyanZ14a, 7780677, 6909475,7298965} can have multiple hidden layers, millions of parameters, billions of operations and require tremendous storage and intense computation resources, making it difficult to realize energy-efficient and high-performance %
solutions. To address this issue, several model compression techniques~\cite{10.5555/2969442.2969588,10.5555/2969239.2969366,DBLP:conf/iclr/MolchanovTKAK17,Jacob_2018_CVPR, DBLP:journals/corr/HanMD15, 8100126}, efficient dataflow techniques~\cite{8114708, 9201450}, and dataflow accelerators~\cite{7551378, 10.1109/ISCA.2016.30, 7738524, 7783723, eyerissv2,  8192478, 8890710, 8702471, 10.1145/3352460.3358291, 8421093, 8983560, 10.1145/3352460.3358275, 9310233, cambricons, nvidia_a100} have been proposed and widely investigated in recent years. 

In particular, exploiting high sparsity in CNNs has emerged as a promising approach for achieving energy-efficient and high-performance CNN solutions~\cite{7551378, 10.1109/ISCA.2016.30, 7738524, 7783723, eyerissv2,  8192478, 8890710, 8702471, 10.1145/3352460.3358291, 8421093, 8983560, 10.1145/3352460.3358275, 9310233, cambricons, nvidia_a100}.
Several sparse CNN (s-CNN) accelerators have recently been proposed to exploit the structured~\cite{cambricons, nvidia_a100} as well as the unstructured sparsity~\cite{7551378, 10.1109/ISCA.2016.30, 7738524, 7783723, eyerissv2,  8192478, 8890710, 8702471, 10.1145/3352460.3358291, 8421093, 8983560, 10.1145/3352460.3358275, 9310233} in both CNN model parameters and its activations. Although this works~\cite {7551378, 10.1109/ISCA.2016.30, 7738524, cambricons, 7783723, eyerissv2,  8192478, 8890710, 8702471, 10.1145/3352460.3358291, 8421093, 8983560, 10.1145/3352460.3358275, 9310233, nvidia_a100} outperform dense CNN accelerators in terms of energy efficiency and performance, they use compressed vector-vector multiplication (VVM) or matrix-vector multiplication (MVM) engines based on traditional dataflow techniques such as weight stationary, input stationary, output stationary, and row stationary techniques~\cite{8114708, 9201450}. When the sparsity of inputs or weights becomes excessively large ($\ge$ 60\%-90\%), previous VVM and MVM accelerators suffer from low utilization issues. For example, when the sparsity is more than 70\%, the utilization of SNAP~\cite{9310233} decreases below 50\%. Furthermore, GoSPA~\cite{gospa} utilization rate falls below 45\% with a sparsity of 90\%, and SCNN~\cite{8192478} falls below 60\% with a sparsity of more than 60\%. As a result, when the sparsity is extremely high, the utilization of these works may be minimal.

\begin{figure}[tb]
\centering
\includegraphics[width=1\columnwidth]{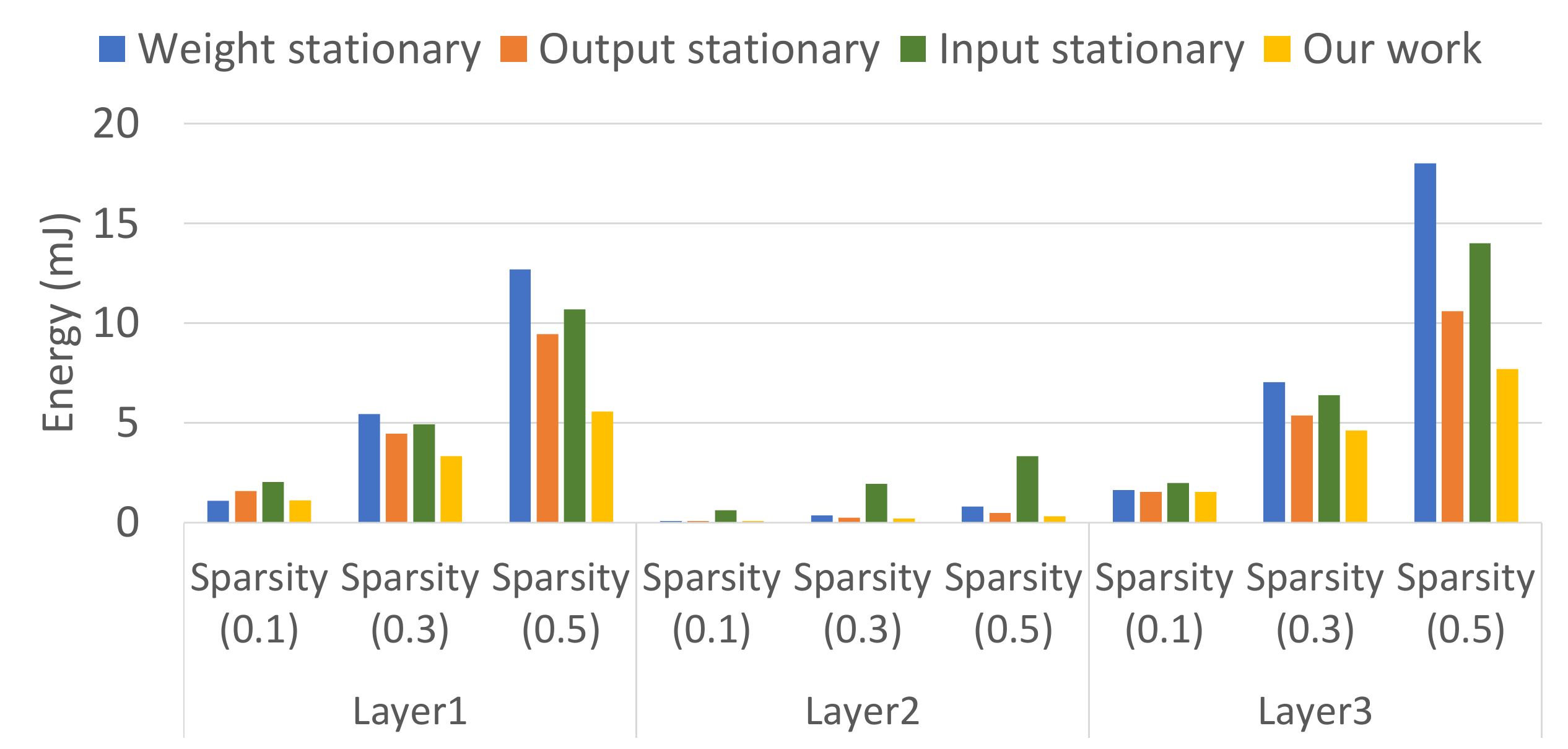}%
\caption{Estimated energy consumption to accelerate three different types of convolution layers (described in Table.~\ref{tab:layers}). These results are estimated using Timeloop~\cite{8695666} and Accelergy~\cite{8942149} tools. Our data show that our event-driven technique consumes less energy than alternate data flows such as weight, output, and input stationary~\cite{8114708, 9201450}. The specifications of three different layers studied are described in Table~\ref{tab:layers}.}
\label{fig:access_energy}%
\end{figure}

\begin{table}[tb]
\centering
\resizebox{1.0\linewidth}{!}{%
\begin{tabular}{|c|c|c|c|c|c|}
\hline
      & \begin{tabular}[c]{@{}c@{}}Input\\  Channels\end{tabular} & \begin{tabular}[c]{@{}c@{}}Output \\ Channels\end{tabular} & \begin{tabular}[c]{@{}c@{}}Input \\ size\end{tabular} & \begin{tabular}[c]{@{}c@{}}Output \\ size\end{tabular} & \begin{tabular}[c]{@{}c@{}}Kernel \\ size\end{tabular} \\ \hline
Layer1 & 256                                                       & 384                                                        & 56                                                    & 56                                                     & 3                                                      \\ \hline
Layer2 & 384                                                       & 256                                                        & 13                                                    & 13                                                     & 3                                                      \\ \hline
Layer3 & 64                                                        & 128                                                        & 224                                                   & 224                                                    & 3                                                      \\ \hline
\end{tabular}
}%
\caption{Example workload used for understanding energy consumption with various dataflow and sparsity options.}
\label{tab:layers}
\end{table}

\begin{figure}[t]
\centering
\includegraphics[width=1\linewidth]{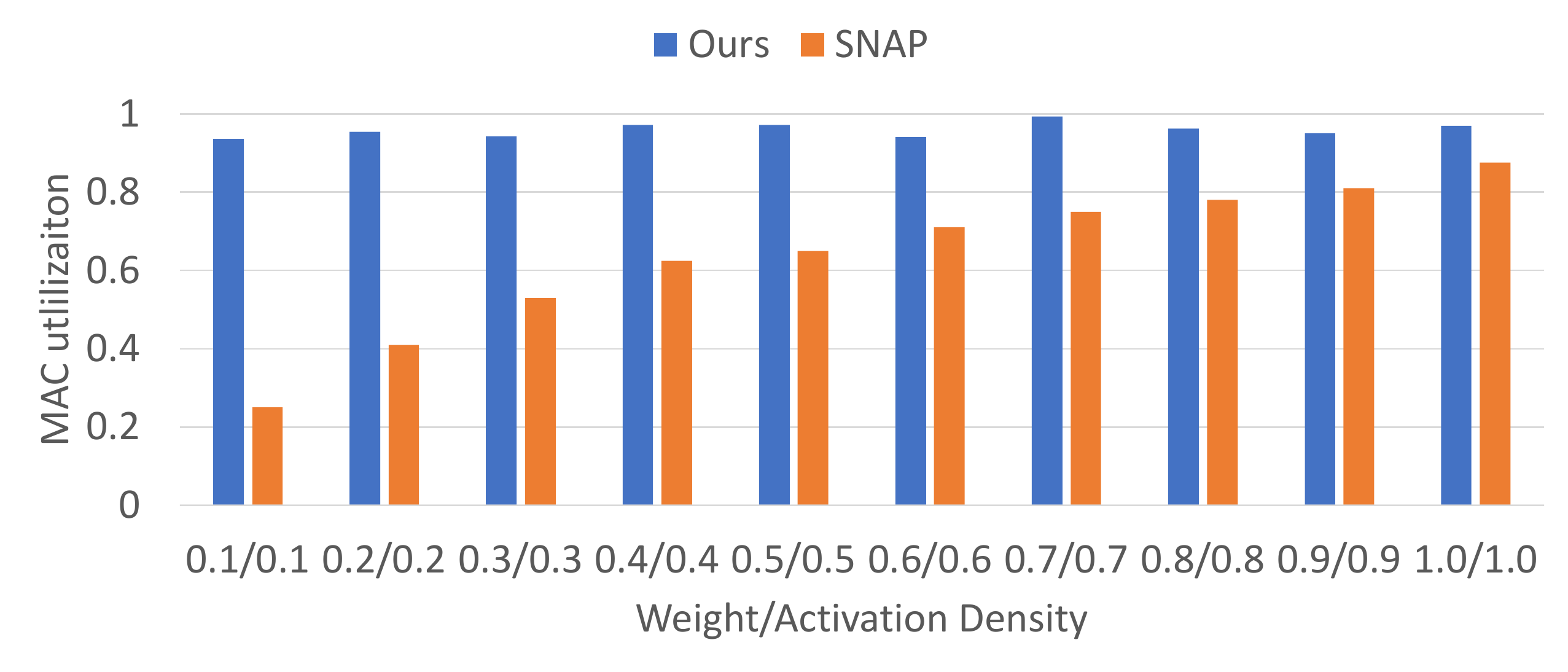}%
\caption{Average multiplier utilization of \nameofwork{} and SNAP~\cite{9310233}. 
}
\label{fig:util}%
\end{figure}

We present an event-driven s-CNN accelerator that maximizes resource efficiency to solve this problem. Each non-zero input channel activation is treated as a single event and processed in an event-driven manner. Memory accesses and multiply-and-accumulate (MAC) operations, in other words, occur only when a processing unit detects an event. When an event needs to be processed, all of the processing elements will do so and update the output neuron value with 100\% resource utilization. If there are no events to process, processing units can enter a low-power idle mode, where the power consumption is reduced by 70\%. When compared to alternative strategies, event-driven processing yields higher resource utilization and a lower overall energy consumption\footnote{Details of the performance and efficiency benefits of our proposed design are discussed in the Evaluation, Section~\ref{sec:evaluation}.}, especially for highly sparse systems (See Figure~\ref{fig:access_energy}). For example, Figure~\ref{fig:util} shows the MAC utilization in our work versus SNAP~\cite{9310233}.

Furthermore, in the state-of-the-art s-CNN accelerators, the sparse input channels or weight filters tend to be encoded in traditional compressed data formats such as compressed sparse row/column (CSR/CSC) and coordinate list (COO)~\cite{10.1109/ISCA.2016.30, 7551378, 8192478,cambricons, eyerissv2, 9310233, gospa}. Therefore, processing units must obtain the entire compressed input channel (CIC) and compressed weight filter (CWF), decode the address and value of non-zero elements and determine the valid pairs of non-zero elements to be multiplied. Reading compressed data in CSR/CSC/COO formats, for example, necessitates the following procedures: 1) gaining access to metadata such as row or column pointers; and 2) determining which column or row indices must be accessed. It must be noted that the number of indices needed varies case by case, and finally, it 3) accesses the non-zero data. As a result of these steps, memory access patterns become complex, irregular and computationally intensive. To address this issue, we propose a novel compressed data storage approach that simplifies memory accesses while minimizing compute overheads and improving overall performance significantly (See Section~\ref{sec:MNF_SECTION}).

Data transfer can also consume more energy than computation, as shown in Table~\ref{tab:energy_per_acc}. DRAM accesses have been found to consume a significant portion of total energy in a wide range of s-CNN accelerators~\cite{7551378, 10.1109/ISCA.2016.30, 7738524, cambricons, 7783723, eyerissv2, 8192478, 8890710, 8702471, 10.1145/3352460.3358291, 8421093, 8983560, 10.1145/3352460.3358275, 9310233, nvidia_a100}. To overcome this issue, we designed an s-CNN accelerator with the fewest possible DRAM accesses and other higher levels of memories, such as global on-chip memories. Our goal, similar to~\cite{7284058}, is to fit all of the CNN parameters on-chip and speed the entire network without accessing DRAMs.

Overall, the main contributions of this work are as follows:
\begin{itemize}
    \item We propose an event-driven sparse CNN accelerator that provides higher efficiency by minimizing memory access to higher-level memories (e.g., DRAMs and global SRAMs) compared to the traditional dataflow techniques. (See Section~\ref{sec:hardware_arch})
    \item We demonstrate how we enable this accelerator by exploiting irregular sparsity in input activations by using a new compressed data storage technique (i.e., event encoding scheme) that allows for direct data access. The proposed event encoding approach increases overall efficiency considerably by simplifying memory accesses. (See Section~\ref{sec:MNF_SECTION})
    \item We present a thorough study of modern dataflow techniques using a variety of sparsity levels and convolution layer shape to demonstrate the advantages of activation-level sparsity. (See Section~\ref{sec:evaluation}) 
    \item Finally, we perform a detailed study and performance comparison with related works. The proposed accelerator surpasses a recent sparse CNN accelerator, GoSPA-R, by 1.46$\times$ on VGG16 and 1.37$\times$ on Alexnet in terms of energy efficiency (frames/J). Furthermore, the proposed accelerator outperforms the GoSPA-R by 1.064$\times$ on VGG16 and 1.33$\times$ on Alexnet in frames processed per second. (See Section~\ref{sec:evaluation})
\end{itemize}

\begin{scriptsize}
\begin{table*}[t!]
\resizebox{1.0\linewidth}{!}{%
\begin{threeparttable}
  \centering
  
  \begin{tabular}{ | *{7}{l|} }
    \hline
    Proposal & SCNN~\cite{8192478}& Cambricon-S~\cite{cambricons} & Eyeriss-v2~\cite{eyerissv2} & SNAP~\cite{9310233}& GoSPA~\cite{gospa} & This work\\
    \hline
    Supported Techniques & Conv & Conv + FC & Conv + FC & Conv + FC & Conv & Conv + FC\\
    \hline
    Data format & CSC & custom weight \& & CSC & COO & CSR & events \\
     & & uncompressed activation & & & &\\
    \hline
    Data precision & 16-bit & variable & 8-bit & 8-bit & 8-bit & 8-bit \\
    \hline
    Architecture & multiple PE & custom design with & Hier Mesh NoC & multi-core & multiple PE & NoC with\\
     & (custom design) & selector modules \& PE & & (custom design) & (custom design) & multicasting\\   
    \hline
    Main memory & DRAM & DRAM & DRAM & DRAM & DRAM & local SRAM\\ 
     & & & &  &  & units\\ 
    \hline
    Data flow & input-stationary & input-stationary & row-stationary & channel-first & input-stationary & event-driven \\
    \hline 
    Strategy to & identifying & indexing & pipelined search & associative & on-the-fly &  \\
    exploit sparsity & non-zero data \& & non-zero data \& & for valid non-zero & index matching & intersection & activation driven \\
     & coord computations  & selector modules & data & & & \\
    \hline 
 
  \end{tabular}
    \caption{Overview of other hardware accelerators for exploiting sparsity. %
  }
  \begin{tablenotes}
        \footnotesize
        \item[] Conv stands for convolutional layer and FC stands for fully connected layer.
        \item[] CSR, COO, CSC stand for compressed sparse row, coordinate format, compressed sparse column, respectively.
        \item[] Note that PE, Hier Mesh, NoC, coord stand for processing element, hierarchical mesh, network-on-chip, coordinate, respectively.
  \end{tablenotes}
  \label{tab:related_work_comp}
\end{threeparttable}
}%
\end{table*}
\end{scriptsize}

\section{Background}
To reduce the overall storage needed to process deep neural network (DNN) workloads, several compression techniques~\cite{10.5555/2969442.2969588,10.5555/2969239.2969366, DBLP:conf/iclr/MolchanovTKAK17,Jacob_2018_CVPR, DBLP:journals/corr/HanMD15, 8100126} have been proposed and widely investigated in recent years. Pruning, in particular, has emerged as a promising compression technique for drastically reducing DNN model size and processing requirements while maintaining accuracy~\cite{10.5555/2969239.2969366, 10.1145/3005348}. Pruning methods can be broadly classified as either unstructured or structured. Unstructured pruning techniques~\cite{10.5555/2969239.2969366} remove neurons and weights with lower importance from wherever they exist. Although these methods achieve very high compression rates by removing nearly 95\% of the model parameters, they produce irregular sparsity, meaning that zeros are distributed randomly across the model parameters~\cite{10.5555/2969239.2969366}. Due to such irregular sparsity, it is difficult to reap the benefits of unstructured pruning methods through traditional hardware solutions such as CPUs, GPUs, TPUs, and ASIC DNN accelerators~\cite{7284058, 10.1145/2541940.2541967, 7011421}.

Structured pruning~\cite{10.1145/3005348}, on the other hand, entails removing a group of weights or neurons or an entire convolutional kernel or filter. Although structured pruning methods have not achieved the compression ratios of unstructured pruning, they produce regular sparsity, which means that zero values appear in groups~\cite{10.1145/3005348}. When the sparsity is regular, it is relatively simple to reap the benefits using traditional hardware solutions~\cite{7284058, 10.1145/2541940.2541967, 7011421}.

In addition to pruning, the rectified linear activation function (ReLU) used in DNN layers reduces DNN processing requirements further. Because all negative activations are treated as zero following ReLU, the number of activations that must be processed after ReLU can be significantly reduced. However, ReLU, like unstructured pruning, produces irregular sparsity ~\cite{pmlr-v119-kurtz20a}. Moreover, the sparsity is input-dependent, which means that the level of sparsity and the location of zeros varies with each input. As a result, unique sparse CNN accelerators are required to capitalize on this sparsity.

\section{Related Work}
\label{sec:related_work}
EIE~\cite{10.1109/ISCA.2016.30} and Cnvlutin~\cite{7551378} were early efforts that used irregular sparsity to enhance CNN throughput and energy efficiency. EIE makes use of both irregular activation and weight sparsity. However, it only supports the fully connected (FC) layers and in many modern image processing tasks CNNs dominate, least amount of computation are completed in the FC layers. Although Cnvlutin supports convolutional (Conv) layers, it only exploits irregular activation sparsity and does multiplications with zero-valued weights.

\subsection{SCNN}

SCNN~\cite{8192478} is the first convolutional neural network (CNN) accelerator that can take advantage of sparsity in activations and weights; however, it only supports the Conv layers. SCNN saves memory by storing activations and weights in a variant of the CSC format. A layer sequencer is used to read the CSC-encoded weights and activations from DRAM and broadcast them to the processing elements (PEs). PEs perform a cartesian product of non-zero weights and activations and then produce the compressed output activations. Unfortunately, the cartesian product method results in many partial sums accumulated and updated in the same output activation address. This can cause write contention and congestion at the output activation memory. Furthermore, to support the Cartesian product technique, SCNN employs an N$\times$N multiplier array, and the utilization rate in SCNN~\cite{8192478} falls below 60\% with a sparsity of more than 60\%.

\subsection{SNAP}
SNAP~\cite{9310233} employs the channel-first dataflow technique, which sorts the weights and activations across multiple channels. An associative index matching (AIM) module is used to determine valid pairs of non-zero weights and activations (i.e., depending on the channel index). The list of valid pairs is subsequently sent to the PEs, which execute the necessary MAC operations. The outputs of MAC operations are then passed to the core reducer and global accumulator to produce the output feature maps (OFMs). While this design reduces the partial accumulated sum write-back traffic, it has a compute core area overhead of 12.5\% for the 16-bit design and 17\% for the 8-bit design. Furthermore, due to the complexity involved in identifying the valid pairs of non-zero data across multiple channels, the utilization rate of the multiplier array implemented drops below 75\% with a sparsity higher than 50\% for both input activation and weights.

\subsection{GoSPA}
GoSPA~\cite{gospa} is one of the most recent works proposed to reduce the area and energy required to discover valid pairings of non-zero weight and activations. It employs a weight-stationary dataflow and suggests an on-the-fly intersection method for determining valid non-zero pairings and a computation reordering strategy for reducing data movement. However, in order to enable both techniques, the GoSPA design employs a dedicated activation processing unit (APU), resulting in 10.6 \% and 14.5 \% space and energy overheads, respectively. Furthermore, GoSPA utilization rate falls below 45\% with a sparsity of 90\%. 

\subsection{Neuromorphic Architectures}

Neuromorphic architectures more closely resemble biological neurons built using spikes and are commonly referred to as spiking neural networks (SNNs). An SNN is a spike-based network that represents information as spikes, and the transmission of the spike is driven by binary events~\cite{p2020spike}. After a neuron computation, if a spike crosses a specified threshold, it will be propagated to the next layer. Otherwise, it will not be propagated. The processing elements are only active when a spike is received and are idle in the absence of spikes. This spike-based communication technique reduces data movement and the number of processing elements needed in SNNs. Several neuromorphic hardware accelerators have been proposed and developed for emulating SNNs.

TrueNorth~\cite{akopyan2015truenorth} is one hardware accelerator designed using an in-memory computing approach. It provides a very low-power density of just 20 milliwatts per square centimetre, but it is only designed for high-spike rates and does not consider sparsity. 
Loihi~\cite{davies2018loihi} is a fully asynchronous neuromorphic chip that can take advantage of sparse activations, support many complex neuron models, and facilitate on-chip training of SNNs with different spiking-time-dependent-plasticity rules. Spinalflow~\cite{narayanan2020spinalflow} explores a novel compressed, time-stamped, and sorted spike input/output representation. %

The event-driven approach in this work is inspired by the SNNs, where each non-zero activation is treated as an event in our work and processes independently in an event-driven manner; similar to that of a spiking SNN.
This means that memory accesses and multiply-and-accumulate (MAC) operations occur only when a processing element detects an event. In addition, one can naturally exploit activation sparsity in CNNs since zero-valued activations can be discarded and will not propagate to the next layer.

\subsection{Related Work Summary}

Table~\ref{tab:related_work_comp} compares the aforementioned works, including our work, in detail. Most of these recent efforts, as indicated in Table~\ref{tab:related_work_comp}, process and store data in compressed formats. In SCNN, zero-operand operations are avoided with the application of a Cartesian product. However, this technique produces many partial sums, which can cause congestion in the data transfer traffic and may cause stalling in the multiplication stage as it waits for the previous results to be stored before computing for the subsequent developments. In Cambricon-S, Eyeriss V2, SNAP and GoSPA, identifying the valid non-zero pairs of input activations and weights increases both area and power consumption.

In contrast to previous research, our proposed \nameofwork{} accelerator employs a straight-forward Multiply-and-Fire (MNF) technique to exploit sparsity in input activations and enable event-driven computation. In the \nameofwork{} architecture, zero-activations are ignored, and only non-zero activations are propagated and processed, as explained in Section~\ref{sec:MNF_SECTION}. Consequently, redundant data transmission, memory accesses, and multiplications are reduced considerably.

\section{Multiply-and-Fire (\nameofwork{}) Approach}
\label{sec:MNF_SECTION}

While existing dataflows attempt to maximize specific types of input or weight data reuse, most fail to take all of them into account. This results in inefficiency when the layer shape or hardware resources vary. In this section, we will introduce our proposed, novel dataflow called Multiply-and-Fire (\nameofwork{}) that aims to accomplish this goal.
In the multiply phase, an input event is used to read the necessary weights and perform multiply-and-accumulate (MAC) operations on the input and weights. A comparator compares the MAC result during the fire phase, which is the output activation, with the predefined threshold. If the output activation exceeds the threshold, a new event is generated using the activation and sent (fired) as a new event to the next layer. If it does not exceed this threshold, the MAC output is not required and discarded (not fired). Since only activations that exceed the threshold are fired as an event to the next layer, the event-driven dataflow naturally exploits activation sparsity and reuse as the computations for the current activation (event) are performed before processing the next one.

\begin{algorithm}[tb]
\textbf{Input:} input: pixel value in the input feature map (IFM)\\
\textbf{Input:} ch\_id: channel index of the IFM\\
\textbf{Input:} start\_neuron: start neuron address\\
\textbf{Input:} start\_weight: start weight address\\

\textbf{Input:} x\_jump: number of steps the filter need to move along x-axis\\
\textbf{Input:} y\_jump: number of steps the filter need to move along y-axis\\
\textbf{Predefined:} nc\_filter: number of columns in weight filter\\
\textbf{Predefined:} nc\_output: number of columns in output feature map (OFM)\\
\textbf{Predefined:} stride of the filter\\

\smallskip
neuron\_addr = start\_neuron \\
weight\_addr = start\_weight \\
\smallskip
\For{y = 0 to y\_jump}{
 \For{x = 0 to x\_jump}{
  neuron[neuron\_addr, ch\_id] += weight[weight\_addr, ch\_id]$\times$input \\
  neuron\_addr += 1 \\
  weight\_addr -= stride \\
 }
 weight\_addr = start\_weight - nc\_filter$\times$($\it{y}$ + 1) $\times$ stride \\
 neuron\_addr = start\_neuron + nc\_output$\times$($\it{y}$ + 1) \\
}

\caption{Algorithm for convolution data processing. %
}
\label{alg:cnn_processing}
\end{algorithm}

In the following subsections, we will describe the algorithms employed in each of the multiply and fire phases.

\begin{figure}[t!]
\centering
\includegraphics[width=.99\columnwidth]{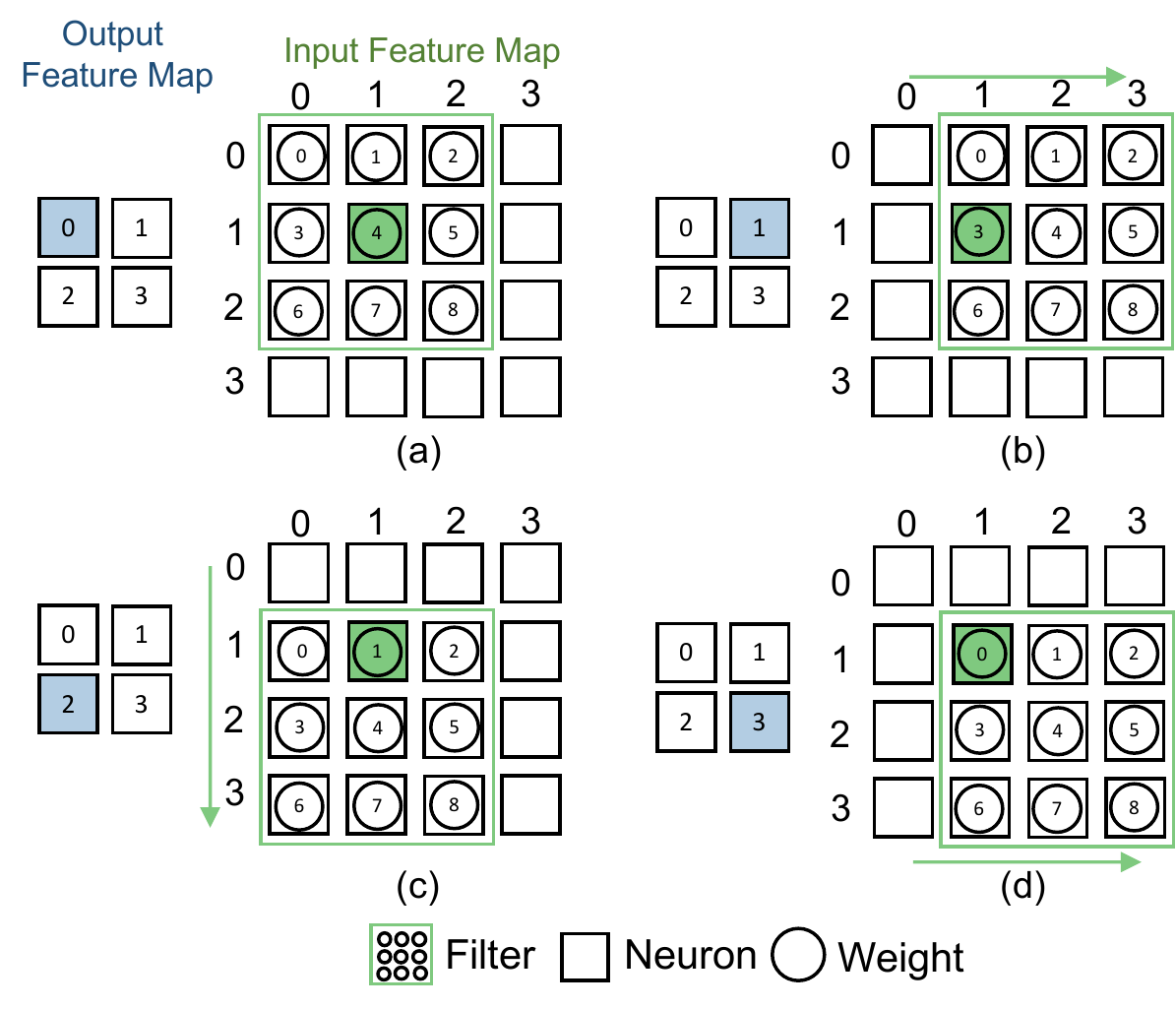}
\caption{The method used by \nameofwork{} to estimate the output feature maps in convolution layers. Given a non-zero input activation at (1,1) position in the input feature map, (a)-(d) shows the step-by-step process used to shift the weight filter and update all the necessary values in the output feature map. 
}
\label{fig:CNN_MNF}
\end{figure}

\subsection{Multiply Phase}
\label{sec:multi_phase}
We consider two types of events to support the acceleration of convolution (Conv) layers and fully connected (FC) layers. The first is for Conv layers, while the second is for FC layers.

\subsubsection{Convolutional layers}
We consider two events to support the acceleration of convolution (Conv) layers and fully connected (FC) layers. The first is for Conv layers, while the second is for FC layers.

Input events for Conv layers have the following information: a) input data, b) channel id, c) start weight address, d) start neuron address, e) x\_jump, and f) y\_jump. Input represents a non-zero pixel value in the input feature map (IFM) of a Conv layer. Channel id represents the channel index of IFM. Start weight address describes the address of the first weight that needs to be multiplied by the input activation value. Start neuron address indicates the address of the first output neuron that needs to receive and accumulate the multiplication output. The x\_jump and y\_jump parameters govern the weight filter's movement over the IFM. For example, the filter will be shifted horizontally by x\_jump steps and vertically by y\_jump steps. Start weight address, start neuron address, x\_jump, and y\_jump are determined by the position of each pixel value in the IFM and stride of the weight filter. %

For example, consider a convolution layer with an IFM size of 4$\times$4 and a weight filter size of $3\times$3. Assume the filter's stride to be one along the x- and y-axes. Suppose the pixel value in the IFM at coordinates (1, 1) is non-zero, such as 100. In that scenario, the input event has the following data: a) input of 100, b) channel id of 0 (assuming only one IFM), c) start weight address of 4, d) start neuron address of 0, e) x\_jump of 1, and f) y\_jump of 1. As stated earlier, the start weight address, start neuron address, x\_jump, and y\_jump depend on the input pixel's location and are all unique for each pixel value in the IFM (See Section~IV).

When a processing element (PE) in \nameofwork{} receives this event, it performs the steps outlined in the Algorithm~\ref{alg:cnn_processing} and detailed below:

\begin{enumerate}[label=(\roman*),noitemsep]
    \item When $\it{x}$ = 0 and $\it{y}$ = 0 in Algorithm~\ref{alg:cnn_processing}, read the weight from memory at location `4' and multiply it by 100. Add the multiplication result to the partial sum of a neuron in the output feature map (OFM) with address `0', as shown in Fig~\ref{fig:CNN_MNF}(a). Next, increase the $\it{x}$ by 1. 

    \item When $\it{x}$ = 1 (i.e., maximum value of $\it{x}$) and $\it{y}$ = 0, subtract the start weight address by stride value of the filter (in this example, 1), and increase the start neuron address by 1. Consequently, the new weight address becomes `3', whereas the new neuron address becomes `1'.
    
    Read the weight from memory at location `3' and multiply it by 100. Add the multiplication result to the partial sum of a neuron in the OFM with address `1', as shown in Fig~\ref{fig:CNN_MNF}(b).
    
    Next, multiply the number of columns in the weight filter (in this example, 3) with ($\it{y}$+1), and stride (in this example, 1). Calculate the new weight address by subtracting the multiplication output (in this example, 3) from the start weight address (in this example, `4'). As a result, the new weight address becomes `1'.  Similarly, perform multiplication of the number of columns in the OFM (in this example, 4), and ($\it{y}$+1). Calculate the new neuron address by adding the multiplication output (in this example, 4) to the starting neuron address (in this example, `0'). 

    \item When $\it{x}$ = 0 and $\it{y}$ = 1 (i.e. the maximum value of $\it{y}$), read the weight from memory at location `1' and multiply it by 100. Add the multiplication result to the partial sum of a neuron in the OFM with address `4', as shown in Fig~\ref{fig:CNN_MNF}(c).
    
    \item When $\it{x}$ = 1 and $\it{y}$ = 1 (i.e., are the maximum values of $\it{x}$ and $\it{y}$), repeat step 2, i.e., subtract the start weight address by stride value of the filter (in this example, 1), and increase the start neuron address by 1. Consequently, the new weight address becomes `0', whereas the new neuron address becomes `5'.
    
    Read the weight from memory at location `0' and multiply it by 100. Add the multiplication result to the partial sum of a neuron in the OFM with address `5', as shown in Fig~\ref{fig:CNN_MNF}(d).

\end{enumerate}

Thus, in this manner, \nameofwork{} performs all necessary multiplications and updates all the required output neuron values in a Conv layer's OFM. The multiplications for a single input event can be processed at once in parallel in the hardware. Next, the approach for handling input events for FCNs is discussed in detail.

\begin{figure}[t!]
\centering
\subfloat[]{%
 \includegraphics[width=0.45\columnwidth]{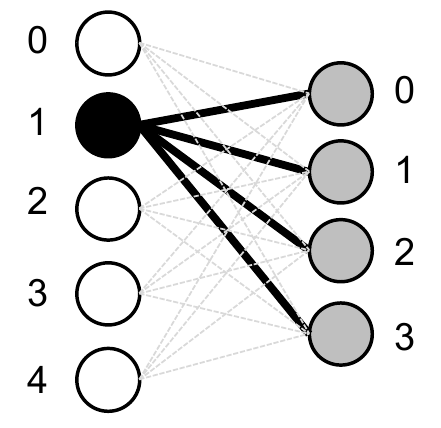}%
 \label{fig:mlp_a}
}
\subfloat[]{%
 \includegraphics[width=0.43\columnwidth]{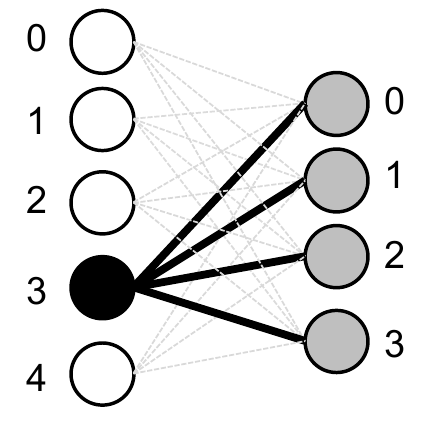}%
 \label{fig:mlp_b}
}
  \caption{The method used by \nameofwork{} to evaluate the outputs of neurons in fully connected networks. (a) if neuron `1' in the first layer has a non-zero input value, all four weights highlighted by the black color are read and multiplied by the input value. Similarly, (b) if neuron `3' in the first layer has a non-zero input value, all four weights highlighted by the black color are read and multiplied by the input value. The multiplication results will subsequently be accumulated at all the neurons in the second layer.}
  \label{fig:MLP_MNF}
\end{figure}

\begin{algorithm}[tb]
\textbf{Input:} input: value of a neuron in the first layer\\
\textbf{Input:} start\_neuron: Address of first neuron in the second layer\\
\textbf{Predefined:} start\_weight: predefined start weight address for input address\\
\textbf{Predefined:} num\_neurons: number of neurons in the second layer\\
\smallskip
neuron\_addr = start\_neuron \\
weight\_addr = start\_weight \\
\smallskip

 \For{x = 0 to num\_neurons-1}{
  neuron[neuron\_addr] += weight[weight\_addr]$\times$input \\
  neuron\_addr += 1 \\
  weight\_addr += 1 \\
 }

\caption{Algorithm for processing input events in fully-connected layers.}
\label{alg:fc_processing}
\end{algorithm}

\subsubsection{Fully connected layers}
The following information is included in input events for FC layers: 1) the input and 2) the address of the neuron. Suppose an FC network has two layers: first and second layers. The input value represents the non-zero activation value of a first-layer input neuron. The neuron address denotes the start address of weights connecting the first layer's input neuron to all neurons in the second layer.

Consider an FC network with two layers: the first with five neurons and the second with four neurons. If neuron '1' in the first layer has a non-zero input activation, such as 100, the input event delivered to the multiply module comprises the following information: a) activation value of 100, and b) neuron address of `1'. 

When the multiply module receives this input event, it performs the steps outlined in the Algorithm~\ref{alg:fc_processing} and detailed below:

\begin{enumerate}[label=(\roman*),noitemsep]
    \item Read all four non-zero weights linked with neuron address `1' and multiply them by input value (in this example, 100). The multiplication results will subsequently be added to all four neuron values in the second layer, as shown in Fig~\ref{fig:mlp_a}.
    
    \item Similarly, if neuron `3' in the first layer has a non-zero input value, all four non-zero weights associated with neuron `3' are read and multiplied by the input value. The multiplication results will then be added to all four neuron outputs in the second layer, as shown in Fig~\ref{fig:mlp_b}.
    
\end{enumerate}

Thus, in this way, \nameofwork{} performs all the required multiplications and updates the output neuron values in FC layers.

\subsection{Fire Module}

After completing the multiply phase, i.e. in the fire module, the values of output neurons in the OFM of a Conv or FC layers are compared with a predefined threshold. If the value of the output neuron exceeds the threshold, it is transformed into an input event and transmitted to the multiply module of the next layer for further processing. On the other hand, if the output value does not surpass the threshold, the fire module ignores the result.

As a result, MNF only fires non-zero output activations to the next layer instead of transferring all activations like the dense model, including zeros. This enables the accelerator to compute data in an event-driven manner and significantly reduces the number of computations in the network and the need to re-compress the activations to the corresponding compression format, leading to lower inference latency and better energy efficiency.

\section{Hardware Architecture and Mapping Technique}
\label{sec:hardware_arch}
This section describes the hardware architecture of \nameofwork{} and the mapping approach used to accelerate convolution neural networks (CNNs) and fully connected networks (FCNs). %

\subsection{Hardware Architecture}

This component's main objective is to enable energy-efficient DNN inference using the Multiply-and-Fire approach presented in Section~\ref{sec:MNF_SECTION}.
The architecture of \nameofwork{} consists of a network-on-chip (NoC) with a mesh topology that connects each PE.
As we will discuss in Section~\ref{mapping_technique}, the PEs can accelerate either a partial layer or an entire layer, depending on the mapping technique. The network itself is based on the OpenSMART NoC design~\cite{opensmart}.

\subsection{Processing Elements}
\label{processing_element_sec}
\begin{figure}[tb]
\centering
\includegraphics[width=0.98\columnwidth]{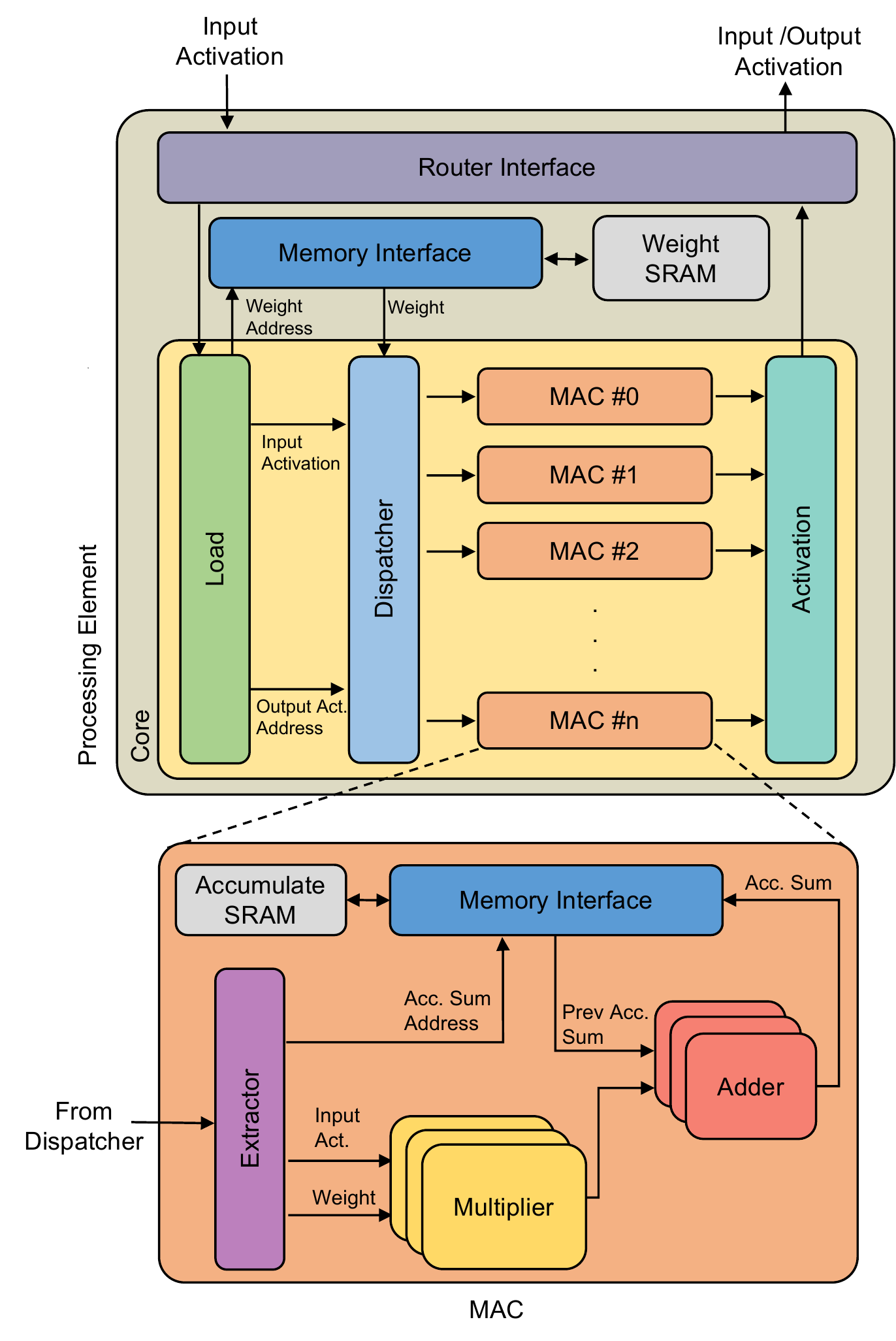}%
\caption{The hardware architecture of a Processing Element (PE) consist of the router interface, core and memory interface modules. Instead of DRAMs as the main storage, each PE has two local SRAMs to store weights and accumulated sums. The core consists of a MAC cluster, load, dispatcher and spiking modules. In the MAC cluster, there are multiple MAC modules and each MAC modules consists of a number of multipliers that are capable of processing multiple MAC operations in parallel.}
\label{fig:arch}%
\end{figure}

Figure~\ref{fig:arch} shows the PE architecture consisting of a router interface, core, memory interface, FIFOs, and two SRAMs to store and process the data locally. The operations of these components are discussed in the following subsections.

\subsubsection{Router interface}
 
Input events are received by the router interface, the gateway into the PE. It is linked to the load and activation modules in the core. It primarily performs three types of operations: delivering the event to the core for processing, forwarding the event to other PEs, and transferring the output results from the core to the next layer (i.e., if two or more PEs need to process the same input). The system can process the information and generate the MAC outputs by sending the event to the core (output activations). The router interface receives the MAC output from the core and sends the result if a MAC output exceeds the threshold. %

On the other hand, the event forwarding mechanism allows the system to employ several PEs to compute the results of multiple channels from one input in parallel.

\subsubsection{SRAMs and Memory Interface}
As seen from Figure~\ref{fig:arch}, each PE contains two SRAMs: a weight and accumulated SRAM. The neural network weights are stored in weight SRAM, whereas the results of MAC operations are stored in accumulated SRAM. The weight SRAM is an ultra-low-leakage single-port SRAM with low power consumption. In contrast, the accumulated SRAM is a two-port SRAM that supports simultaneous read and write operations. In addition, both of the SRAMs are clock gated to improve efficiency. Each SRAM has its memory interface, which receives memory read and write requests from the core, and the MAC clusters then send them to SRAMs.

\subsubsection{Core}

Figure~\ref{fig:arch} depicts the design of the core. It comprises a load module, a dispatcher module, a MAC cluster for parallel computation, and an activation module. %
The architecture is built on a decoupled access/execute approach~\cite{smith1982Decoupled}, which allows the modules to operate independently as long as the necessary data is available.

When the MAC module is stalled, for example, the load module can continue loading data to serve the dispatcher modules, and the dispatcher module can still generate new results to be processed by the MAC module. The decoupled access/execute micro-architecture results in increased compute and memory access overlap, which reduces computational delay. Instead of connecting the modules directly, circular FIFOs are used as an interface between them to enable this result.

If an input event arrives via the router interface, the core reads the data, configures its control unit, and broadcasts the value to all modules. Each module has two operations based on actual input events and end-of-data events. End-of-data events indicate that all input events have been processed and that the PEs are ready to send the results to be processed by the next layer.

\textbf{Load Module:} The load module is responsible for reading the data required to execute the multiply and fire phases introduced in Section~\ref{sec:MNF_SECTION}. 

The load module begins by decoding the input events received from the router interface. Based on the decoded channel index and start weight address, it generates the actual weight address in weight SRAM and then sends a read request to the weight SRAM interface. In addition, it generates a list of output neuron addresses that the current input will be updated based on the decoded start neuron address. The load module packs the addresses in the correct order and sends them and the input activation value to the dispatcher module.

When an end-of-data event is received after all input data has been received, the load module does not compute any additional addresses or send any new read requests; it only forwards the event to the dispatcher module. In both the operations, the load module determines the type of input received (actual input event or end-of-data event), decodes the event and forwards the information to the dispatcher module. 

\textbf{Dispatcher Module:}
The dispatcher module has two main tasks. First, it groups the input activation value with its weight and the corresponding output neuron address. Second, based on the neuron address, it sends this data to the respective MAC module to be processed and accumulated.

\begin{figure}[tb]
\centering
\includegraphics[width=.92\columnwidth]{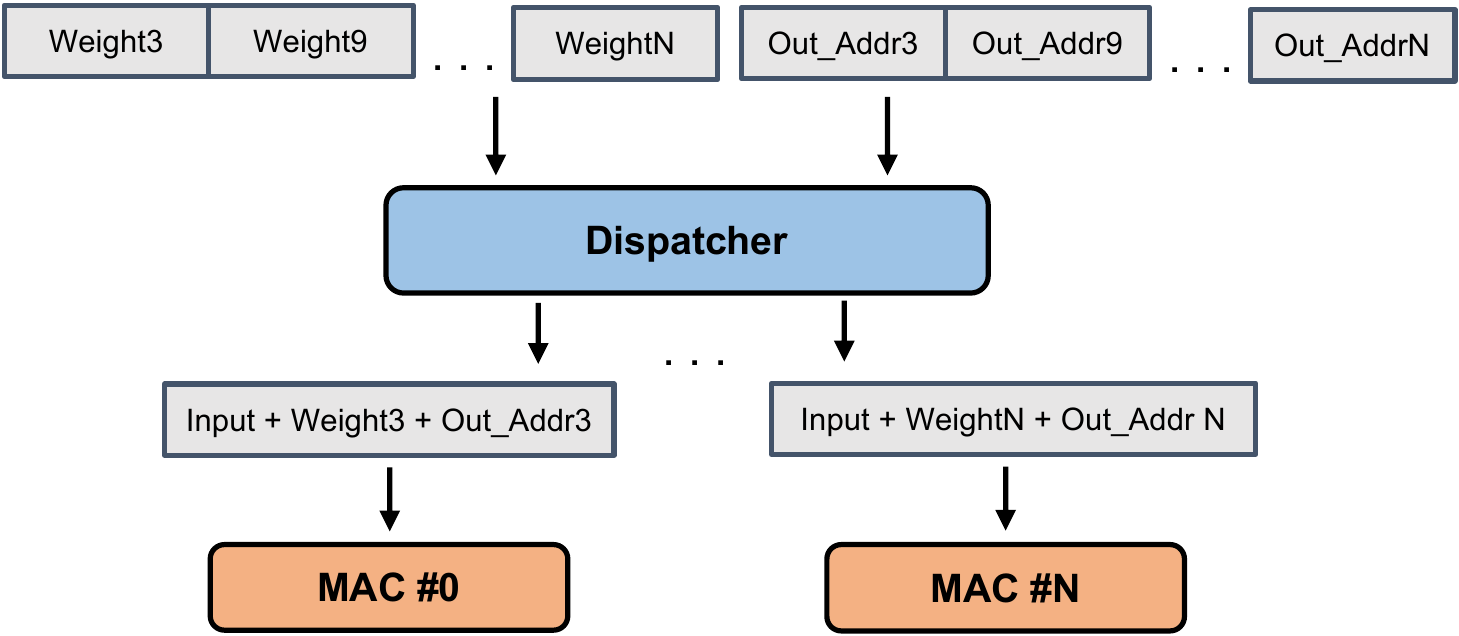}%
\caption{ The dispatcher module groups input activation, weight and corresponding the output neuron address received, then send the data to the respectively MAC module based on the neuron address. This module ensures that the groups of data are sent to the correct MAC module to be processed.}
\label{fig:dispatcher}%
\end{figure}

The core is designed with a MAC cluster with multiple MAC modules to enable parallel computation; a single read from weight SRAM retrieves a vector of weights. Similar to how we process the weights, multiple output neuron addresses are packed together in a vector by the load module. When these data items reach the dispatcher, individual weights and neuron addresses are extracted from the vectors then grouped accordingly. The process can be seen in Figure \ref{fig:dispatcher}, where the weight and output address data is grouped with the input and sent to the corresponding MAC unit to be processed. In our design, the number of weights retrieved in a single read equals the number of multipliers implemented. This allows the multiplication between weights retrieved and the input in parallel in a single cycle.

\begin{figure*}[tb]
\centering
\includegraphics[width=0.78\linewidth]{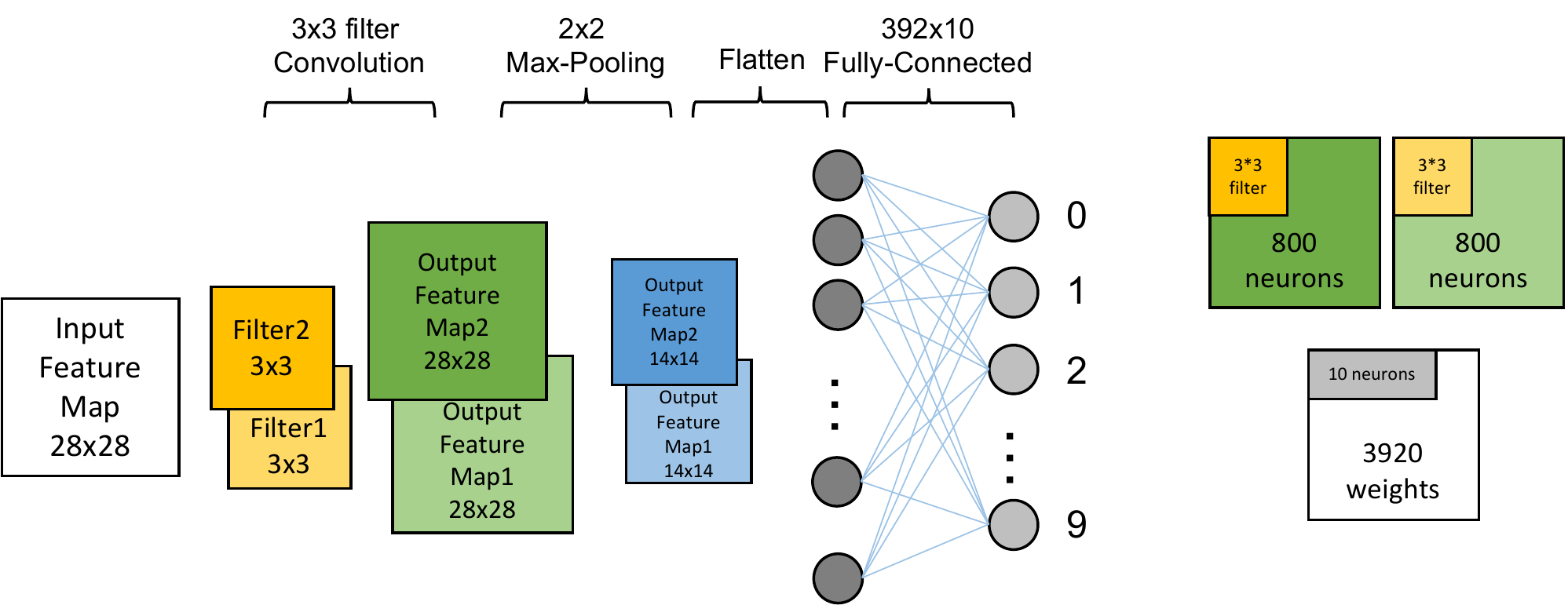}%
\caption{An illustration of the mapping technique developed for accelerating Convolutional Neural Networks (CNNs) on \nameofwork{}. Consider a simple CNN architecture consists of one convolution layer with two 3$\times$3 filters, one max pooling layer and one fully connected layer. As shown, for a memory capacity of 800 neurons and 9000 weights, three PEs (i.e., are represented as squares on the right) are needed to accelerate the CNN layer. }
\label{fig:mapping_example}%
\end{figure*}

\textbf{MAC Cluster:}
The MAC cluster is responsible for executing the multiply and accumulation computation of the 8-bit fixed point input activations and weights. It also interfaces with the accumulated SRAM directly through the memory interface module, as seen in Figure \ref{fig:arch}, to retrieve and store the 32-bit partial sum. The partial sums are immediately accumulated with the local SRAM to the designated output neurons.

When it receives data from the dispatcher module, the MAC module performs the following computations:
\begin{enumerate}
    \item Split the data into input, weight and output neuron address.
    \item Multiply the input with the weight.
    \item Retrieve the previous accumulated result from the neuron address.
    \item Add the previous accumulated sum with the newly computed result.
    \item Store the new accumulated sum back to the same neuron address.
\end{enumerate}

After processing each of the input events, the MAC module receives the end-of-data event and performs the following computations: 

\begin{enumerate}
    \item Retrieves the accumulated sum from the SRAM  
    \item Quantize the sum to an 8-bit integer value following the technique proposed in~\cite{Jacob_2018_CVPR}
    \item Transfer the 8-bit sum to the activation module
\end{enumerate}

In the MAC cluster, there are multiple MAC modules and each MAC modules is capable of processing multiple MACs in parallel. For example, for the VGG16 network with filter size 3$\times$3, we can implement 9 MAC modules each with 3 multipliers. In total, 3 3$\times$3 which is equivalent to 27 MACs can be computed in parallel.

\textbf{Activation Module:}
The activation module is responsible for performing max-pooling and ReLU operations. It is connected to the MAC cluster. Since max-pooling and ReLU only occur after all MAC computations are done, it is only activated when the system receives the end-of-data event and idle the other time.

For layers where max-pooling operations are not required, the activation module will sequentially read results from the MAC cluster and perform the ReLU operation. This operation is completed by comparing the output activation value with a predefined threshold (typically zero). 

If the output activation read is larger than the threshold of ReLU, the output is propagated to the next layer. The additional information (channel id, start weight address, start neuron address, x\_jump, and y\_jump) required to form the input event of the valid result is also computed. This information and the result are combined to generate a new event, transferred to the router interface and then to the next layer. 

For layers where the max-pooling operation is required, the activation module will read several results to perform the max-pool operation from the MAC to the activation module. It finds the maximum value among the output activations to perform the ReLU operation. Consider a case with a $2\times2$ max-pool operation: 4 results will be read from the MAC cluster, with the exact MAC modules depending on the neuron address. 

\begin{figure*}[tb]
\centering
\subfloat[AlexNet]{%
  \includegraphics[width=0.4118\linewidth]{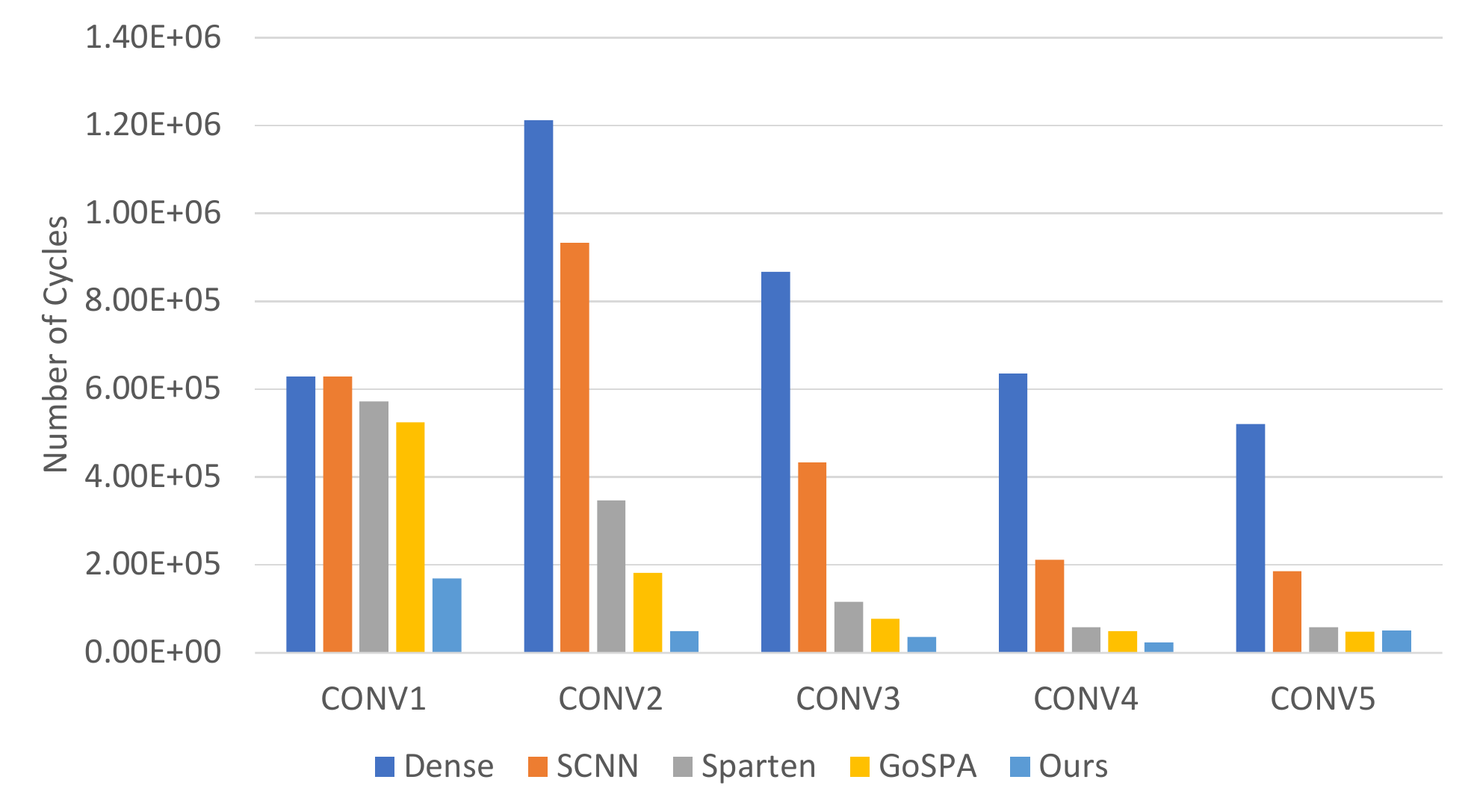}%
  \label{fig:layer-alexnet}%
}
\subfloat[VGG16]{%
  \includegraphics[width=0.5882\linewidth]{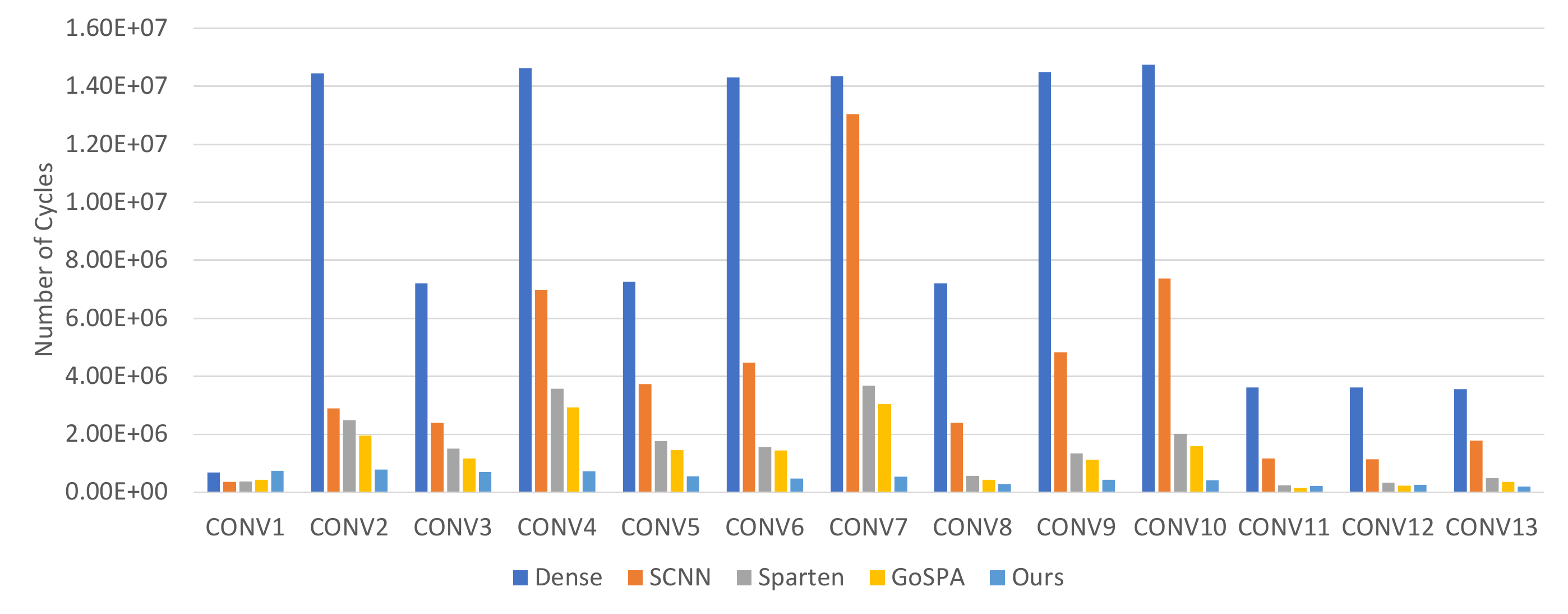}%
  \label{fig:layer-vgg}%
}

\caption{The number of cycles required comparison of Dense, SCNN, Sparten, GoSPA, and \nameofwork{} on AlexNet and VGG16.
}
\label{fig:layer_level_cycles}
\end{figure*}

\begin{table}[tb]
\centering
\begin{tabular}{lS[table-format=3.2]} 
\hline
Component                    & {Size}                                                                        \\ \hline
Number of PEs               & 11                                                                          \\
MAC Cluster Size            & 9                                                                           \\
Multipiers per PE           & 27                                                                          \\
Weight SRAM per PE (KB)     & 691.2                                                                       \\
Accumulate SRAM per PE (KB) & 67.5                                                                        \\
Frequency (MHz)             & 200                                                                         \\
Bit Precision               & {\begin{tabular}[c]{@{}r@{}}Weight/Acc: 8 bits \\ Psum: 32 bits\end{tabular}} \\ \hline
\end{tabular}
  \caption{\nameofwork{} Specifications}
  \label{tab:hardware_spec}
\end{table}

\subsection{Mapping Technique}
\label{mapping_technique}

This section describes the neural network mapping technique used in this work. The mapping technique proposed enables CNNs to be run on the proposed hardware efficiently and helps eliminate the need for memory transactions between different PEs, with an overview visualized in Figure~\ref{fig:mapping_example}. 

The mapping technique calculates the number of PEs needed given a specific neural network configuration. The number of PEs required to accelerate a Conv or FC layer is determined by the storage capacity of the PE's SRAMs. For instance, the minimum number of PEs ($C\_PEs$) required to generate an output feature map of a Conv layer with size $w \times h$ can be estimated using Equation~\ref{eq:minpecnn}. 

\begin{equation}
\label{eq:minpecnn}
 C\_PEs= max(\frac{w \times h}{N},\frac{k \times k \times c}{W})
\end{equation}

Where $N$ and $W$ are the maximum number of neurons and weights that a PE can store in its SRAMs, the size of the weight filter is represented by $k \times k$, and the number of weight filters is represented by $c$. The $C\_PEs$ can be arranged in a NoC grid with dimensions as large as $\lceil \sqrt{C\_PEs} \rceil \times \lceil \sqrt{C\_PEs} \rceil$.
For example, consider a Conv layer with $28\times28$ input feature map (IFM) with padding of one and two $3\times3$ weight filters. This Conv layer generates two output feature maps (OFMs) of size $28\times28$. Assuming that the accumulated and weight SRAM of each PE can only store the data of 800 neurons and 9000 weights, respectively, a minimum of two PEs are required to create those two OFMs according to the Equation~\ref{eq:minpecnn}. In the design, the storage size of the accumulated SRAM should be big enough to store the neurons of an entire channel to eliminate the complexity of spreading the computation of a channel to multiple PEs. Since the two OFM are mapped across multiple PEs, the pixel values in the IFM must be shared across the PEs.

Similarly, the minimum number of PEs ($F\_PEs$) required to accelerate an $m\times{}n$ FC layer can be estimated using Equation~\ref{eq:minpemlp}. For example, if we consider an FCN with two layers: 1568 neurons in the input layer and 128 neurons in the output layer, the 128 neurons in the output layer must be mapped across a minimum of 23 PEs, and weight SRAM capacity now dictates the size in the Equation~\ref{eq:minpemlp}. Again, because the neurons of the output layer are mapped over many PEs, the values of the input neurons must be shared across the PEs.

As we process the network layer by layer, the PEs used in the Conv layer computation are reused by the FC layer. In addition to the computational PEs described in \ref{processing_element_sec}, a storage PE with minimum logic and large storage capacity is needed to receive and store the outputs from the PEs. These outputs will be sent as input events to computational PEs for subsequent layer processing.

\begin{equation}
\label{eq:minpemlp}
 F\_PEs = max(\frac{n}{N},\frac{m \times n}{W})
\end{equation}

\begin{table*}[tb]
\centering
\resizebox{.9\linewidth}{!}{%
\begin{tabular}{cc*{6}S[table-format=3.2]|S[table-format=3.2]}
\toprule
\multicolumn{2}{c}{Design}                       & {GoSPA-R}   & {GoSPA}     & {Eyeriss}   & {EyerissV2} & {NullHop}  & {Ours}        & {Ours(22nm)} \\
\midrule
\multicolumn{2}{c}{Bit Width}                    & 8         & 16        & 16        & 8         & 16       & 8           & 8          \\
\multicolumn{2}{c}{\#MACs}                       & 128       & 128       & 168       & 128       & 128      & 297         & 297        \\
\multicolumn{2}{c}{TOPS}                         & {6.40e+10}  & {6.40e+10}  & {7.80e+10}  & {5.94e+10}  & {6.40e+10} & {5.94e+10}    & {5.94e+10}   \\
\multicolumn{2}{c}{Frequency (MHz)}              & 500       & 500       & 464       & 464       & 500      & 200         & 200        \\
\multicolumn{2}{c}{Accelerator Area (mm2)} & 1.8   & 2.7   & 2.3   & 4.1   & 6.3  & 8.3   & 6.5     \\
\multicolumn{2}{c}{Total Area (mm2)*} & 78.0   & 78.9   & 78.5   & 80.3   & 82.5  & 122.2   & 96.0     \\
\multirow{2}{*}{Frames/S}         & VGG16        & 29.7      & 29.7      & 1.9       & {-}         & 13.7     & 31.6        & 31.6       \\
                                  & AlexNet      & 460.3     & 460.3     & 89.6      & 795.0     & {-}        & 612.1       & 612.1     \\
\multirow{2}{*}{Power (mW)}       & VGG16        & 277.0     & 429.4     & 269.0       & {-}         & 257.0    & 200.5       & 171.4     \\
                                  & AlexNet      & 290.0     & 445.0     & 336.0     & 1075.0    & {-}        & 280.5       & 239.7     \\
\multirow{2}{*}{Frames/J}         & VGG16        & 107.3     & 69.2      & 6.9       & {-}         & 53.3     & 157.6       & 184.4    \\
                                  & AlexNet      & 1587.0    & 1034.0    & 266.7     & 739.5     & {-}        & 2182.2      & 2553.1  \\ \bottomrule
\end{tabular}
}%

\begin{tablenotes}
        \footnotesize
        \item *Assumed that other designs are using Hynix DDR5-6400 8Gb chip to store weights, input activations and output feature maps.
    \end{tablenotes}
\caption{Performance comparisons among different sparse CNN ASIC designs (Power and Frames/J are scaled to 28nm except for the last column).
}
\label{tab:compare_with_others}
\end{table*}

\section{Evaluation}
\label{sec:evaluation}

\subsection{Experimental Methodology}
To compare our work and other accelerators, we use ImageNet data as the input to evaluate the performance of AlexNet and VGG-16. The evaluated sparse models are trained and pruned using Pytorch, maintaining a similar accuracy to dense models. For pruned VGG16 and AlexNet, they can achieve 76.04\% and 52.75\% on ImageNet compared with dense models, which are 75.77\% and 52.72\%, respectively. After pruning, the overall weight sparsity ratio is 49.9\% and 59.6\% for AlexNet and VGG16, respectively. 

Our accelerator is synthesized using Synopsys Design Compiler version P-2019.03, targeting a 22 nm technology node with 11 PE. Gate-level simulations are performed using Synopsys VCS-MX K-2015.09, and power analysis was performed with Synopsys PrimePower version P-2019.03. All the simulations and analyses are performed at 200 MHz. A detailed hardware specification is shown in Table \ref{tab:hardware_spec}.

\subsection{Performance Analysis}
For the comparison of SCNN-Dense, SCNN, Sparten, GoSPA, and \nameofwork{}, we evaluated the number of cycles needed to run VGG16. We estimated the number of cycles on AlexNet by using the same hardware configuration (shown in Table \ref{tab:hardware_spec}. The evaluated sparse models are trained and pruned using Pytorch and maintain a similar accuracy compared to the dense models. After pruning, the overall weight density ratio is 49.9\% and 59.6\% for AlexNet and VGG16, respectively. Figure \ref{fig:layer_level_cycles} shows the number of cycles required to process different sparse CNN models. From the figures, it can be seen that compared with SCNN-Dense, SCNN, Sparten, and GoSPA because \nameofwork{} has a better dataflow and high efficient dataflow based hardware design, \nameofwork{} reduces the number of cycles needed to compute VGG16 by 19.0$\times$, 8.31$\times$, 3.15$\times$, and 2.57$\times$ for SCNN-Dense, SCNN, Sparten and GoSPA, respectively.

On AlexNet, \nameofwork{} can achieve 11.82$\times$, 7.32$\times$, 3.51$\times$, and 2.68$\times$ improvement in cycle count over SCNN-Dense, SCNN, Sparten, and GoSPA, respectively. Notice that \nameofwork{} is slightly inferior to other designs on the first layer of VGG16. This is because the first layer of VGG16 is very dense. For example, GoSPA can use large FIFOs to handle the very dense layers, but the large FIFOs will become an overhead for the rest sparse layers. We expect \nameofwork{} would achieve higher performance than other accelerators on the dense layers if extra hardware resources are added as GoSPA does. 

The high MACs utilization help \nameofwork{} outperform other designs. \nameofwork{} has very high utilization of multipliers. Figure \ref{fig:util} shows that the utilization of multipliers under various weight/activation densities is always close to 100\% (the utilization is slightly different between various density levels because the number of channels is not always a multiple of the number of MACs available). Compared with SNAP~\cite{9310233} whose utilization is influenced by the sparsity significantly, \nameofwork{} has a stable and higher MACs utilization on all sparsity levels.

\subsection{Comparison with other sparse CNN accelerators}
We also compare our work with other sparse CNN accelerators like GoSPA, the Eyeriss series, and NullHop. All results are scaled to the 28nm technology node for a fair comparison. According to the quantization algorithm used in our design, the bit width is the same with GoSPA-R and Eyeriss V2. Based on the 8-bit design, we synthesize our method with a 200\,MHz operating frequency.
As shown in Table \ref{tab:compare_with_others}, our work outperforms other sparse CNN ASIC designs both in speed and energy efficiency. In terms of energy efficiency (in frames/J), \nameofwork{} beats Eyeriss by 22.8x and 8.18x on VGG16 and AlexNet, respectively. \nameofwork{} can also achieve at least 1.46$\times$ and 1.37$\times$ higher energy efficiency than GoSPA-R on VGG16 and AlexNet, respectively. \nameofwork{} achieves a 16.99$\times$ speed up on VGG16 compared with Eyeriss, and it can achieve at least a 1.06x speedup compared with GoSPA-R. Compared to the power consumption of these designs, \nameofwork{} achieves at least a 1.29$\times$ and 1.03$\times$ improvement compared to VGG16 and AlexNet, respectively.

\subsection{Power Breakdown}
\begin{figure}[tb]
\centering
\includegraphics[width=1\columnwidth]{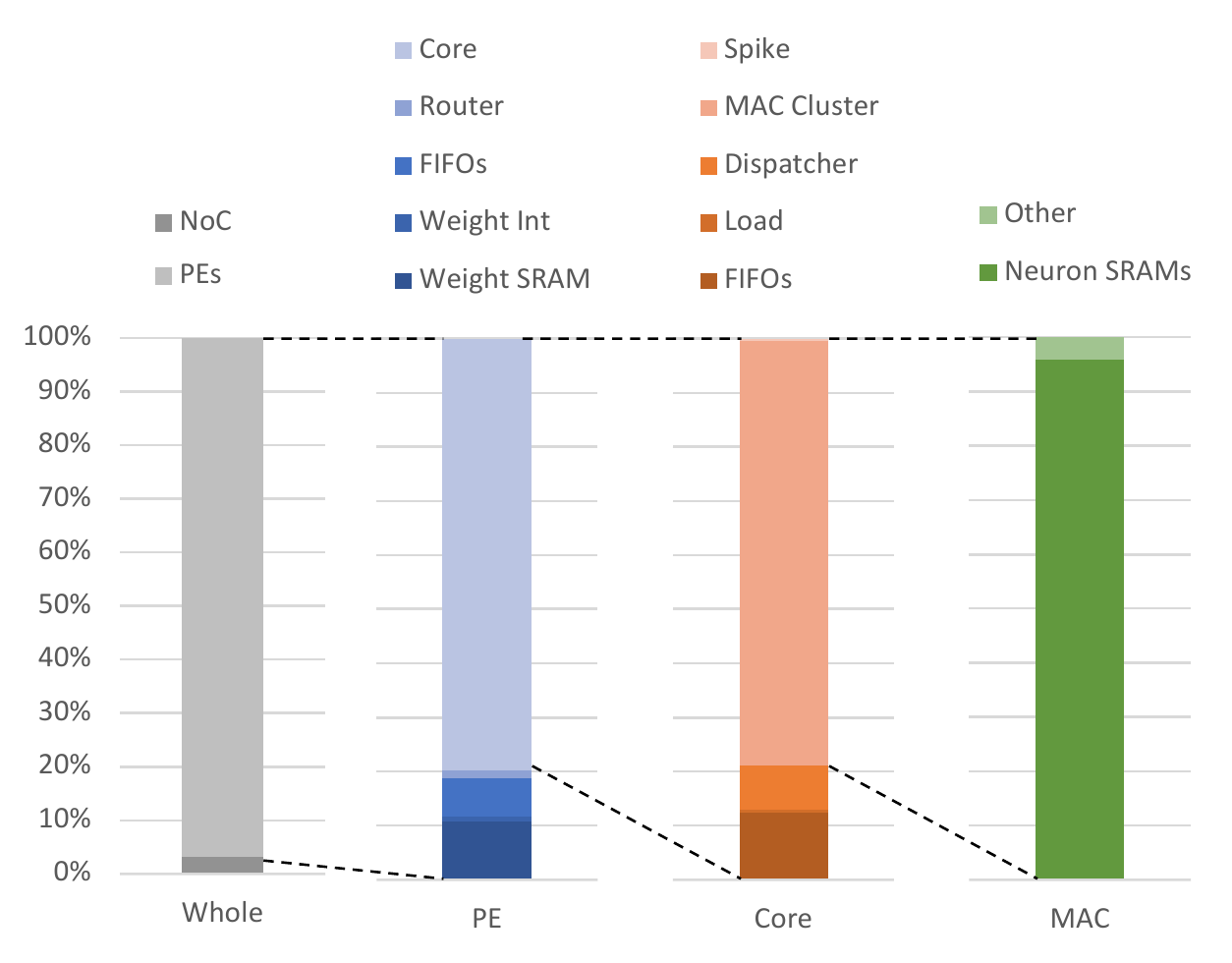}%
\caption{Power Distribution Breakdown of the \nameofwork{} Hardware Accelerator.}
\label{fig:pe_power}%
\end{figure}

The power consumption of the processing element and core of the \nameofwork{} accelerator is evaluated. Figure~\ref{fig:pe_power} 
breaks down the power consumption of each major hardware component inside the processing element shown listed in Figure~\ref{fig:arch}. We find that core consumes around 80\% of the total power used by the PE.
Furthermore, Figure~\ref{fig:pe_power} also shows the power breakdown inside the core. The MAC cluster takes most of the power of the core design. Step into the MAC clusters, the neuron SRAMs take more than 90\% Power.

\subsection{\nameofwork{} Memory Access Efficiency}

In this section, we compare \nameofwork{} with weight stationary, output stationary, and input stationary dataflow models. We modeled these dataflows using Timeloop~\cite{8695666} and Accelergy~\cite{8942149} and estimate the number of memory accesses and energy consumed for accessing different levels of memory. Although our work requires additional accesses to the local buffer and register file than input, output, and weight stationary dataflow types, we can see that, given the energy cost values from Table~\ref{tab:energy_per_acc}, that the overall energy cost 
is lower with our technique %
(See Figure~\ref{fig:access_energy}). 
For different workloads (Table~\ref{tab:layers}) and sparsity, \nameofwork{} can achieve lower energy consumption due to the focus on accessing local buffers instead of larger caches and DRAM.

\begin{table}[tb]

  \centering
  \resizebox{1.0\linewidth}{!}{%
  \begin{tabular}{ | *{5}{l|} }
\hline
      & \multicolumn{2}{|c|}{Other dataflows~\cite{8695666, 8942149}}  &  \multicolumn{2}{|c|}{Our work}   \\ \hline
     & Datawidth & Energy & Datawidth & Energy \\ \hline
DRAM & 64 bits & 512 pJ & 32 bits & 256 pJ \\ \hline
SRAM   & 64 bits & 74 pJ & 32 bits   & 3.87 pJ\\ \hline
PE Buffer  & 16 bits & 1.59 pJ & 216 / 32 bits & 12.35 / 3.87 pJ \\ \hline
Register   & 16*3bits & 0.97*3 pJ & 8*3 bits & 0.018*3 pJ \\ \hline
  \end{tabular}
  }
  \caption{Memory access energy consuption in our work vs others \cite{8695666, 8942149}. 
}
  \label{tab:energy_per_acc}
\end{table}

\section{Conclusion}

This work presents a novel activation-sparsity-driven hardware accelerator for AI inference workloads. By combining our unique per-activation dataflow and highly-parallel computation, this work has achieved state-of-the-art efficiency and performance for a set of standard AI benchmarks to the best of our knowledge. We feel that parallel, on-chip processing can help achieve high performance and efficiency for common ANN tasks, including many upcoming workloads where sparsity is now commonplace.

\bibliographystyle{IEEEtranS}
\bibliography{refs}

\end{document}